\documentclass[conference]{IEEEtran}
\usepackage{graphicx}
\usepackage{tabularx}
\usepackage{cite}
\usepackage{amsmath}
\usepackage{textcomp}
\usepackage{enumitem}
\usepackage{booktabs}
\usepackage{multirow}
\usepackage{siunitx}
\usepackage{balance}
\usepackage{xcolor}

\ifCLASSINFOpdf
\else
\fi

\begin{document}

\title{Decision Tree Based Hardware Power Monitoring for Run Time Dynamic Power Management in FPGA}

\author{\IEEEauthorblockN{Zhe Lin\IEEEauthorrefmark{1},
Wei Zhang\IEEEauthorrefmark{1} and
Sharad Sinha\IEEEauthorrefmark{2}}
\IEEEauthorblockA{\IEEEauthorrefmark{1}Department of Electronic and Computer Engineering, Hong Kong University of Science and Technology, Hong Kong}
\IEEEauthorblockA{\IEEEauthorrefmark{2}School of Computer Engineering, Nanyang Technological University, Singapore\\
\{zlinaf, wei.zhang\}@ust.hk, sharad\_sinha@ieee.org}}

\maketitle

\begin{abstract}
Fine-grained runtime power management techniques could be promising solutions for power reduction. Therefore, it is essential to establish accurate power monitoring schemes to obtain dynamic power variation in a short period (i.e., tens or hundreds of clock cycles). In this paper, we leverage a decision-tree-based power modeling approach to establish fine-grained hardware power monitoring on FPGA platforms. A generic and complete design flow is developed to implement the decision tree power model which is capable of precisely estimating dynamic power in a fine-grained manner. A flexible architecture of the hardware power monitoring is proposed, which can be instrumented in any RTL design for runtime power estimation, dispensing with the need for extra power measurement devices. Experimental results of applying the proposed model to benchmarks with different resource types reveal an average error up to 4\% for dynamic power estimation. Moreover, the overheads of area, power and performance incurred by the power monitoring circuitry are extremely low. Finally, we apply our power monitoring technique to the power management using phase shedding with an on-chip multi-phase regulator as a proof of concept and the results demonstrate 14\% efficiency enhancement for the power supply of the FPGA internal logic.
\end{abstract}

\IEEEpeerreviewmaketitle

\section{Introduction}
\label{sec:intro}
With the growth of capacity and complexity of field programmable gate arrays (FPGAs), and the increasing usage of FPGAs in data centers, the reduction of the FPGA power consumption is becoming an important issue. To alleviate this problem, FPGA vendors provide gate-level power analysis tools for power estimation, e.g., Xpower analyzer (XPA) from Xilinx and Powerplay from Altera. These tools are fully aware of the internal hardware implementation and they can provide accurate power estimation during design time, thus helping designers to tune their circuits using power optimization techniques during development periods. Nevertheless, there is a burgeoning interest in applying runtime power management techniques, e.g., dynamic voltage frequency scaling (DVFS)~\cite{pm16} and task scheduling techniques~\cite{Losch16 }. Such runtime strategies make it a necessity to be aware of the runtime dynamic power consumption for the applications running on FPGA. One common approach to obtain runtime power of FPGAs is to use power measurement circuits. This method, however, suffers from two major disadvantages: (1) it requires additional board area to integrate the power measurement devices; and (2) the power detection period is long, in the granularity of milliseconds~\cite{vrscale}. As dynamic power management approaches, such as DVFS and current control using on-chip regulators, become feasible and
promising~\cite{keller}, dynamic power monitoring in the granularity of tens or hundreds of cycles is required so as to support fine-grained power reduction.

Within this context, our goal is to establish a fine-grained, accurate yet light-weight dynamic power monitoring scheme for FPGA-based designs, leveraging the state-of-the-art machine learning theory. We first propose a generic and complete design flow capable of capturing the key features necessary for power modeling on FPGA and generating samples for power modeling. Based on that, we exploit a decision-tree-based power modeling method and develop an in-situ supporting architecture which can be efficiently integrated into the RTL designs with extremely low overheads of performance, power consumption and resource utilization. The power estimation period can be as fine-grained as tens of clock cycles, facilitating more possibilities of power management strategies. The experiments reveal that the proposed decision-tree-based power model exhibits salient improvement in both accuracy and area overheads in comparison to traditional linear regression methods. Furthermore, we propose model ensemble and runtime phase shedding with an on-chip multi-phase regulator to tackle practical problems using our hardware power monitoring scheme.

In general, our contributions can be summarized as follows:

\begin{itemize}
\item A platform-independent and complete synthesis flow to extract features, generate activity traces and power traces as samples for power model establishment.
\item A runtime dynamic power monitoring scheme based on decision tree learning theory, with complete analysis from feature selection to model optimization.
\item A light-weight and in-situ hardware realization of the power monitoring scheme, including activity counters and an area-efficient memory-based decision tree with small overheads of power, performance and resource.
\item A model ensemble strategy: we experimentally quantify the error of aggregating the pre-trained power models as an ensemble model and shed light on its viability for library-based and IP-based designs.
\item A proof of concept for fine-grained phase shedding using an on-chip multi-phase voltage regulator for FPGA internal logic.
\end{itemize}

The rest of this paper is organized as follows. Section~\ref{sec:related} gives a general discussion about modern power modeling approaches. Section~\ref{sec:flow} and Section~\ref{sec:dtree} demonstrate the complete design flow and Section~\ref{sec:wrapper} elaborates the monitoring hardware. Experimental results are discussed in Section~\ref{sec:exp} and finally, we conclude the paper in Section~\ref{sec:con}.

\section{Related Work}
\label{sec:related}
In the literature, researchers have shown great interests in dynamic power modeling for FPGAs, as it is one of the main challenges for FPGA-based designs. Studies about FPGA power modeling have been conducted at different abstraction levels. In the work~\cite{li03}, a low level abstraction model was developed by capturing average switching power based on switched-level macromodels for LUTs and registers. Likewise, the work~\cite{najoua} derived power models from basic operators, e.g., adders and multipliers. These low level models are specialized in their targeted FPGA devices and thereby make them difficult to migrate to other FPGA families. In contrast, power models at high abstraction level are more promising to be generalized in a shorter time, using a set of key signals without the necessity of looking into lower level details.

In the recent work~\cite{laksh,najem,kapow}, linear regressions have been employed for high level power models. Work~\cite{laksh} established the regression model using IO toggle rate and resource utilization in a log-arithmetic format. It attempted to formulate a generic model suitable for all applications running on FPGA. However, the conditions different from the trained model such as the variation in IO port size will significantly degrade the estimation accuracy. Work~\cite{najem,kapow} extracted the toggle rate of a small set of internal signals and built the model in embedded processors, resulting in the LUT resource overhead of 7\% in~\cite{najem} and 9\% in~\cite{kapow} for their tested applications, as well as around 5\% CPU time for both of these two work. Furthermore, as studied in~\cite{lee15,bogl}, power behaviors of complex arithmetic units are generally non-linear. Hence, the linear model exhibits intrinsic restriction in accuracy enhancement when non-linear power patterns increase with the growing sample size, which is known as the underfitting problem in machine learning theory. In light of this problem, the objective of our work is to leverage a decision tree learning model with the ability to adaptively learn different power patterns from samples captured under various situations. To the best of our knowledge, our work proposes the first approach to establish non-linear dynamic power models
in FPGA using the state-of-the-art machine learning theory. Furthermore, the overheads of area, power and performance incurred by our proposed power monitoring is trivial. The resulted area overhead of the power monitor is negligible in comparison with prior work~\cite{najem,kapow}.

\section{Automatic synthesis flow}
\label{sec:flow}
\begin{figure}[ht]
\begin{center}
\vspace{-3mm}
\includegraphics[width=\linewidth]{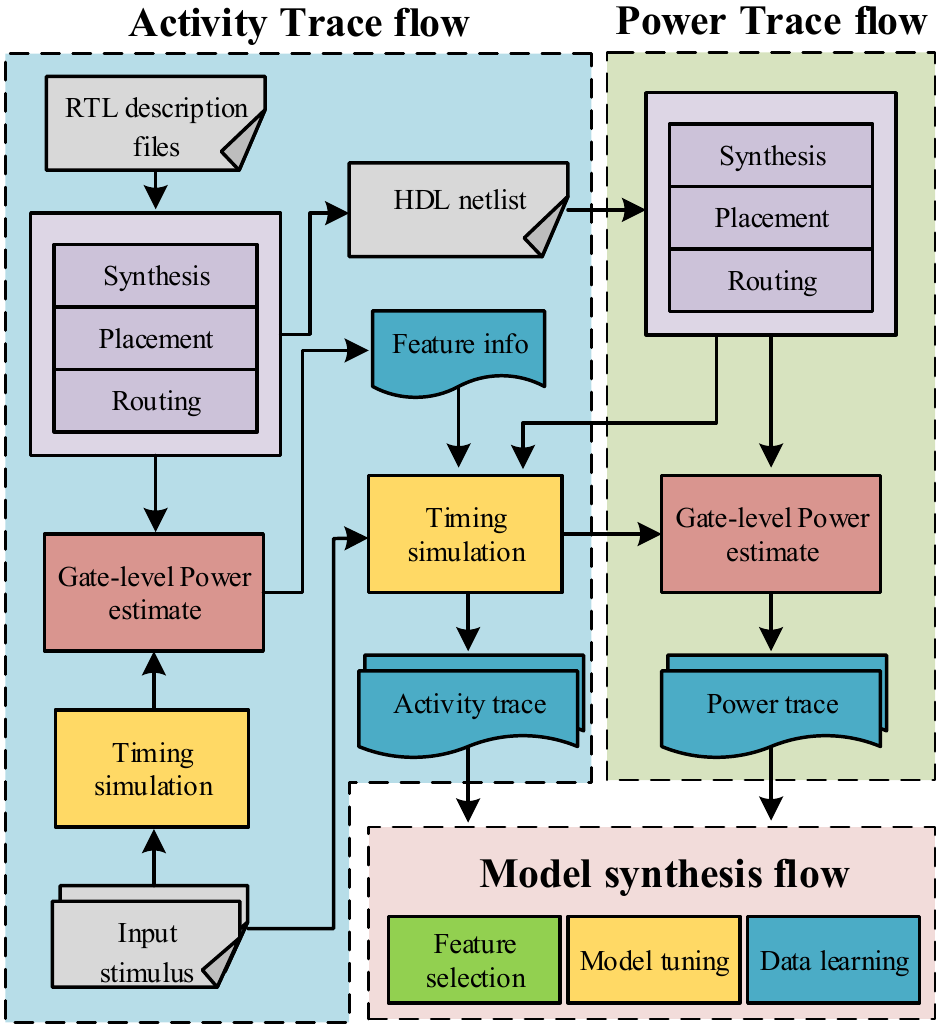}
\caption{Automatic synthesis flow.}
\label{fig:flow}
\vspace{-6mm}
\end{center}
\end{figure}

Provided the original register transfer level (RTL) designs, we propose a complete automatic synthesis flow from power and activity trace generation to the decision tree model establishment as shown in Fig.~\ref{fig:flow}. The flow can be decomposed into three sub-flows: (1) activity trace flow (ATF); (2) power trace flow (PTF); and (3) model synthesis flow (MSF). An activity trace is defined as the switching activities of a set of signals in the design in an estimation period, whereas a power trace is the power value of the target design in an estimation period. Note that our implementation targets Xilinx tool chain and Modelsim, but the generic methodology is applicable to other vendor tools.

\subsection{Activity trace flow (ATF)}
Dynamic power is introduced by signal transitions which dissipate power by charging and discharging the load capacitors. With the improvement on leakage power control in recent devices, the main consideration for power management is more relevant to dynamic power which is the main focus of our work.
The equation of dynamic power dissipation is summarized in Equation~(\ref{eq:dyn}), which demonstrates the relationship between switching activity $\alpha_i$, capacitance $C_i$ on the net $i$, supply voltage $V_{dd}$ and operating frequency $f$. Here voltage and frequency are pre-set constants while capacitance is determined by the device and switching activities totally depend on the runtime execution of the application.
\begin{equation}
\label{eq:dyn}
P_{dyn}=\sum_{i \in N}\alpha_iC_iV_{dd}^2f
\end{equation}

In order to generalize the dynamic power models at high level, a set of key signals are captured and we monitor their switching activities formulated in Equation~(\ref{eq:act}), where $\boldsymbol{a}(\cdot)$ is the activity function which returns the difference of signal transition counts $\boldsymbol{s}(\cdot)$  on the signal set $\boldsymbol{sig}$ over the estimation period from $t_{start}$ to $t_{end}$. For a large design containing millions of nets, it is vital to identify a subset of discriminative and informative nets that are strongly indicative of the power. The number of transitions of an identified signal in an estimation period serves as a feature for the power model synthesis.
\begin{equation}
\label{eq:act}
\boldsymbol{a}(\boldsymbol{sig})=\boldsymbol{s}(\boldsymbol{sig},t_{end})-\boldsymbol{s}(\boldsymbol{sig},t_{start})
\end{equation}

In order to identify the most indicative signals, we first run vector-based timing simulation with randomly generated input vectors and then use Xpower to export the sorting of the signal activities. The simulation is supposed to run for a sufficiently long time (i.e., orders of magnitude larger than an estimation period) to cover as many situations as possible. We select a number of candidate signals with highest activities, and then perform feature selection in MSF. The feature selection is performed to filter out the redundant signals showing repetitive behaviors and leave a smaller subset of signals with discriminative features across different input vectors, as discussed in Section~\ref{subsec:fs}. Note that the identified signals are the lower level nets in the HDL netlist exported by command \emph{write\_verilog} after placement and routing. The HDL netlist is basically composed of primitives (e.g., instances of LUTs or DSPs) and connections. The main advantages of using the HDL netlist are two-fold: it enables us to extract the identified signals, and it preserves the original mapping when the activity counters are added afterwards for implementing the model in hardware for runtime monitoring. Finally, timing simulations are conducted using Modelsim to generate \emph{.saif} files for power synthesis in PTF, with the identified features recorded in activity traces simultaneously.

\subsection{Power trace flow (PTF)}
We go through synthesis and implementation again using the netlist in the PTF flow. After that, we apply the \emph{.saif} files derived from ATF to the Xpower analyzer to derive the corresponding power traces for different estimation periods. To attain high confidence power values, we need to set proper constraints for IO toggling and all the clock specifications. In addition, more than 25\% of the signal toggling should be covered in the \emph{.saif} files~\cite{ug910}. The static power is dependent on the ambient temperature. As we target Virtex-7 series FPGAs with Xilinx high performance low power (HPL) technology and we note that the static power for Xilinx HPL technology is small in magnitude and shows small variation in a wide range of temperatures, i.e., within 1 W for the power range from -40 to 60 degree centigrade~\cite{hpl}, which also conforms with our experimental results. Hence, we conduct our experiments under the ambient temperature of 25 degree centigrade to approximate the static power whereas we mainly investigate the dynamic power.

\subsection{Model synthesis flow (MSF)}
After we collect a sufficient amount of activity traces and the corresponding power traces, our anchor is to develop a runtime dynamic power monitoring scheme based on decision tree learning algorithm. As introduced later in Section~\ref{subsec:dt}, the decision tree can eventually be decomposed into a series of if-then-else rules and intrinsically this is suitable to be implemented in an area-efficient way in hardware with a series of comparators. Moreover, the decision tree caters for models with different complexities, by tuning the inherent parameters (e.g., maximum depth, minimum number of samples to split a node). It also renders high adaptability to improve itself when more samples are provided, as indicated by the learning curves in Section~\ref{subsec:assess}. In comparison, the traditionally employed linear regression model shows an insufficiency in the ability to acclimate itself to further increase accuracy when more training samples are available. Nevertheless, the decision tree is prone to the overfitting problem. In order to build an accurate decision tree regression model, we systematically present the essential steps about the decision tree model establishment in Section~\ref{sec:dtree}.

\section{Decision tree model establishment}
\label{sec:dtree}
\subsection{Background of decision tree}
\label{subsec:dt}
\begin{figure}[ht]
\begin{center}
\vspace{-2mm}
\includegraphics[width=0.8\linewidth]{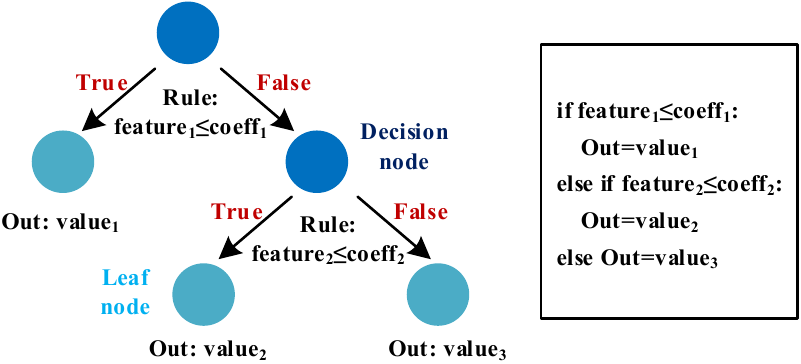}
\vspace{-2mm}
\caption{Graphical and textual representation of decision tree regression.}
\label{fig:back}
\vspace{-5mm}
\end{center}
\end{figure}

The decision tree is a nonparametric hierarchical model for supervised learning which learns the samples in the form of a tree structure. At each step of the tree building process, a decision rule is generated, in which a coefficient is used to compare with the target feature. According to the result of the comparison, the sample set is split into two homogeneous subsets, as shown in Fig.~\ref{fig:back}. Basically a tree node can be categorized as: (1) decision/internal node: a node where the contained subdivision of sample set can be further split by a feature; and (2) leaf/terminal node: a node associated with an output which can not be further split. The decision nodes represent the decision rules whereas the leaf nodes provide the classification or regression results. We use the CART~\cite{cart} algorithm to develop decision tree models.

\subsection{Feature selection}
\label{subsec:fs}
In ATF, we have already screened out a set of candidate signals with high activities. Before we develop the power models, we apply feature selection to preserve only the signals containing informative features across different input vectors in different estimation periods. Feature selection can circumvent the overfitting problem and reduce the resource overhead induced by activity counters used to monitor signals. We leverage recursive feature elimination algorithm to identify required features, which has been successfully applied to other domains~\cite{ravish}. In each iteration, the decision tree model is trained and the feature importance values are computed, according to the node impurity at different splits, which is also known as the Gini importance~\cite{cart} in the decision tree. The features with least importance are eliminated and the remaining features are used to re-train the decision tree model and update feature importance. The insignificant features are recursively pruned away and a subset of 20\% key features is remained for the final model establishment. This maximum subset size is empirically determined based on the observation that the final decision tree will use 10\% to 20\% of the features as we extract one hundred features for each application.

\subsection{Model tuning and training}
The problem of learning an optimal decision tree is known to be NP-complete under several aspects of optimality. To boost the performance of our target model, we seek to tune a set of essential hyper-parameters which largely determine the accuracy of training, validation and testing. We focus on the parameters defined in Table~\ref{table:param}. The most influential parameter for the decision tree is the maximum depth which controls how deep a tree can be formed. A deep tree means a high degree of model complexity while also makes it prone to overfitting on data. On the contrary, a shallow tree often fails to learn from the samples well. The tree depth should be tuned in terms of different applications to strike a good balance between training and testing accuracy. In addition, the other three parameters are supposed to be coordinated with the tree depth and be tuned to circumvent the overfitting problem~\cite{cart}.

We experimentally determine the sets of parameters suitable for our benchmarks: \{3, 4, 5, 6, 7, 8\} for maximum depth, \{5, 10, 15, 20\} for minimum split sample and minimum leaf sample, and \{0.001, 0.01, 0.02, 0.03, 0.04, 0.05\} for minimum leaf impurity. To determine the best set of hyper-parameters that fits the data well, k-fold cross validation approach~\cite{rodri} is employed. We adopt ten-fold cross validation method: the training set is further divided into ten subsets. A decision tree model is trained using nine of them and the model is evaluated with a score using the left subset as the validation set. This procedure is repeated ten times with a different validation set each time and we take the average to get the overall score of the set of parameters, as an quantitative indicator for performance. We try every combination of the hyper-parameters and conduct cross validation to quantify the performance of each model. The set of hyper-parameters with highest score will be deployed to develop the model. After that, we use both the training set and the validation set to train the model in the model developing process. Finally, we assess the model with the test set.
\begin{table}[ht]
\centering
\small
\vspace{-2mm}
\caption{Hyper-parameters for Decision tree tuning.}
\vspace{-2mm}
\label{table:param}
\begin{tabular}{p{.2\linewidth}|p{.65\linewidth}}
    \toprule
\textbf{Name}&\textbf{Description}\\
    \midrule
Maximum depth & The maximum depth that a tree can grow to. \\ \midrule
Minimum split sample & The minimum number of samples used to split a decision node.\\  \midrule
Minimum leaf sample & The minimum number of samples necessary to determine a leaf node.\\ \midrule
Minimum leaf impurity & The minimum percentage of samples giving different output at a leaf node.\\
    \bottomrule
\end{tabular}
\vspace{-3mm}
\end{table}

\subsection{Model ensemble strategy}
\begin{figure}[ht]
\begin{center}
\includegraphics[width=0.9\linewidth]{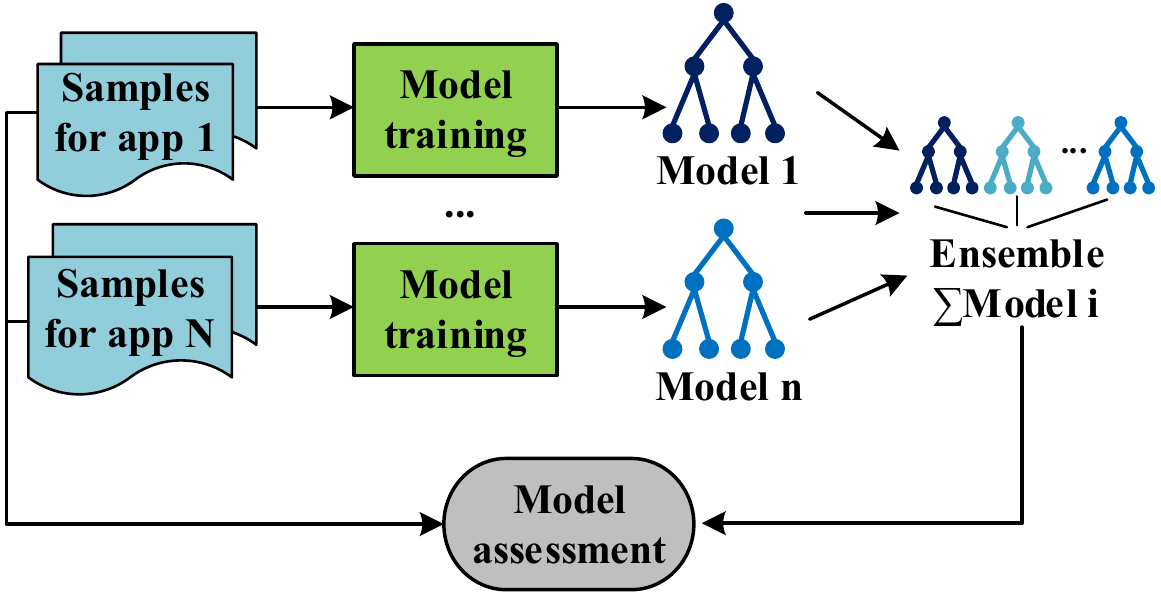}
\vspace{-2mm}
\caption{Model ensemble flow.}
\label{fig:aggre}
\vspace{-4mm}
\end{center}
\end{figure}

A large-scale hardware design traditionally requires the collaborative participation of a group of designers, each of whom focuses on an independent part of the complete design. In light of this, we exploit the practical usage of our model when different power models are developed independently for different parts of the design and eventually they are aggregated together and behave as an ensemble power model. The model ensemble flow is shown in Fig.~\ref{fig:aggre}. Since the dynamic power for different components are derived from different decision tree models, we directly add up the dynamic power estimated from the corresponding decision tree models in order to obtain the overall power consumption. Experiment in Section VI-D shows that the additional error is as small as 1.2\% compared to the re-trained decision tree model.

The model ensemble facilitates the model reuse: a power model trained for a specific function can be saved in a library and be integrated into an aggregated model directly when needed. It is applicable for library-based and IP-based designs with encapsulated power models. With the increasing usage of FPGAs in data centers to accelerate multiple functions loaded at runtime, such a fast and accurate ensemble method will be particularly useful and necessary.

\section{Hardware wrapper architecture}
\label{sec:wrapper}
We propose an area-efficient hardware wrapper to realize the proposed power monitoring in FPGA. The hardware wrapper can be decomposed into two parts: activity counter and decision tree regression engine.
\subsection{Activity counter}
\begin{figure}[ht]
\begin{center}
\includegraphics[width=\linewidth]{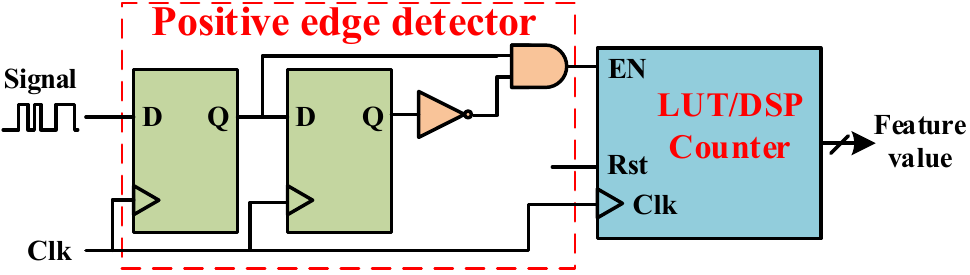}
\caption{Activity counter architecture.}
\label{fig:actcnt}
\vspace{-7mm}
\end{center}
\end{figure}

The activity counter design is shown in Fig.~\ref{fig:actcnt}. A positive edge detector firstly identifies the positive edges of the input signal. Its output is valid for a single cycle, acting as the enable signal to the counter. We propose two realizations of the counter design: the LUT-based counter and the DSP-based counter. The LUT-based counter only utilizes LUT resource whereas the DSP-based counter is implemented using the primitive \emph{COUNTER\_LOAD\_MACRO}, as the instantiation of the dynamic loading up counter occupying one DSP48 unit with a maximum data width of 48 bits. These two counter templates render the flexibility for developers who are aware of the application resource utilization. Since we count the positive edges of selected signals, the number of clock cycles in an estimation period implies the upper bound of the signal activity, which can be referred to set the maximum bit width for counters. The counters are reset at the beginning of every detection cycle by the decision tree regression engine.

\subsection{Decision tree regression engine}
\label{subsec:dtreereg}
\begin{figure}[ht]
\begin{center}
\includegraphics[width=\linewidth]{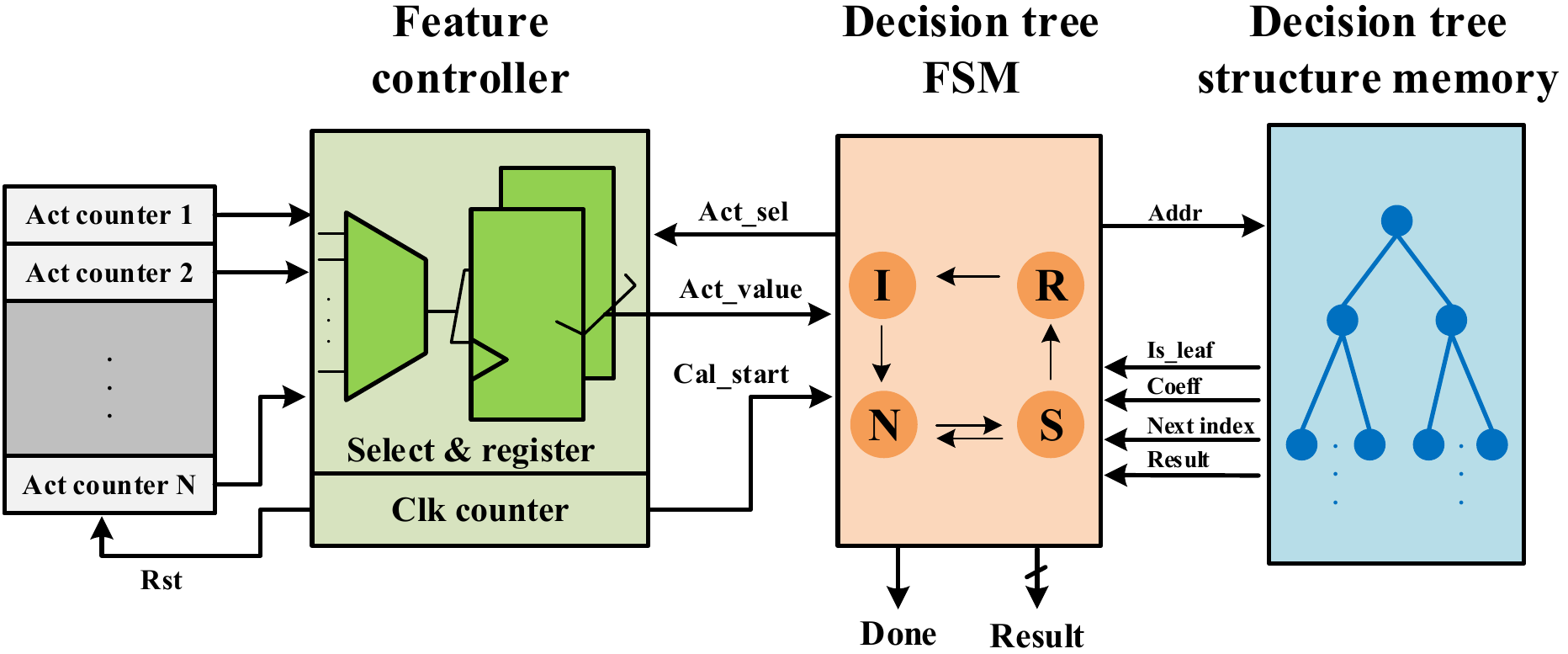}
\vspace{-7mm}
\caption{Decision tree regression engine.}
\label{fig:dtree}
\vspace{-8mm}
\end{center}
\end{figure}

We propose a memory-based decision tree regressor as shown in Fig.~\ref{fig:dtree}. There are some studies on decision tree implementation in hardware~\cite{saqib} to maximize the throughput of decision tree computation. Nevertheless, our objective is different from prior work in that the prime consideration is not throughput in our solution. Instead, our decision-tree-based monitoring scheme only operates once in an estimation period and thereby we target at reducing power and resource overheads at the first place. In our solution, the decision tree structure is completely preserved in a memory element. Additional peripheral control units are incorporated to orchestrate the activity counter controlling, tree node decoding and branch deciding. To summarize, the complete hardware structure in our proposed solution can be further divided into three subsystems: (1) a feature controller;  (2) a decision tree finite state machine (FSM); and (3) a decision tree structure memory.

\begin{figure}[ht]
\begin{center}
\includegraphics[width=\linewidth]{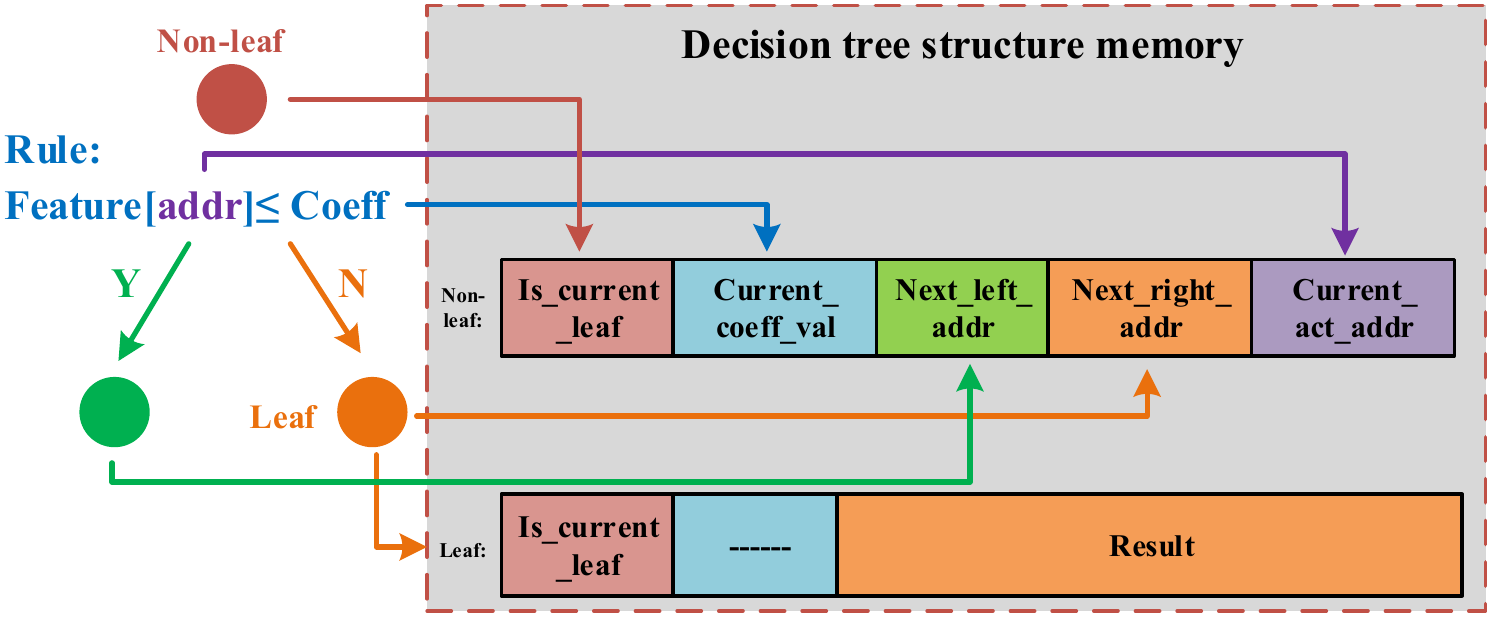}
\vspace{-6mm}
\caption{Decision tree memory structure.}
\label{fig:dtreeram}
\vspace{-5mm}
\end{center}
\end{figure}

To achieve periodic power estimate, the feature controller buffers feature values from activity counters, invokes the FSM to operating states and periodically resets the activity counters. We use a user-specified parameter to set the period of the power estimation. The decision tree FSM has four states: idle (I), node reading (N), stalling (S) and result outputting (R). The FSM starts execution by transferring the state from idle to node reading. In the node reading state, the FSM fetches the node information and tree structure from the memory and completes the if-then-else branch decision by comparing the addressed feature with the decoded tree node coefficient. Stalling state will operate together with the node reading state to ensure the correctness of memory reading. The execution finally terminates by transferring to result outputting state which gives the final result and sets the indication signal when the tree leaf has been reached. The decision tree structure memory is the fundamental component which preserves the complete tree structure and the feature addresses for rule decisions, as shown in Fig.~\ref{fig:dtreeram}. The memory uses the block memory on FPGA.

The maximum execution time for a single invocation is $2n+1$ cycles where $n$ is the maximum depth of the tree. Note that the decision tree structure is completely preserved in the structure memory, meaning that the proposed hardware wrapper is generally applicable to all decision tree types varying in depth or different pruning structures. In addition, the features are the number of signal transitions in an estimation period and therefore, they are unsigned integer numbers. Correspondingly, the coefficients can also be revised in an unsigned integer format when doing the rule decision, without loss of precision. Following this observation, no floating point operations are required for the hardware wrapper design, which contributes to the high area efficiency of its implementation. Furthermore, our proposed hardware wrappers are applied to HDL netlists and the HDL netlists are lower level designs comprising primitives and connections to describe the mapping of the RTL designs after placement and routing. As a result, the integration of the monitoring circuits preserves the RTL mapping.

\section{Experimental results}
\label{sec:exp}
Our proposed activity trace flow and power trace flow are implemented in Vivado 2016.4 and we utilize Modelsim SE 10.3 for the generation of the \emph{.saif} files. The targeted FPGA platform is Virtex-7 XC7V2000tflg1925-1. The model synthesis flow is developed based on the Scikit-learn 0.18.1~\cite{scikit} machine learning toolbox. We applied our methodology to develop the decision-tree-based power models and respectively build hardware wrappers for several benchmarks in Chstone~\cite{chstone}, Polybench~\cite{polybench} and Machsuite~\cite{machsuite}. These benchmark suites are C-based benchmarks and we derive the synthesizable Verilog version using Vivado HLS 2016.4.

\subsection{Model assessment}
\label{subsec:assess}
\begin{figure*}[ht]
\begin{center}
\includegraphics[width=5.9cm]{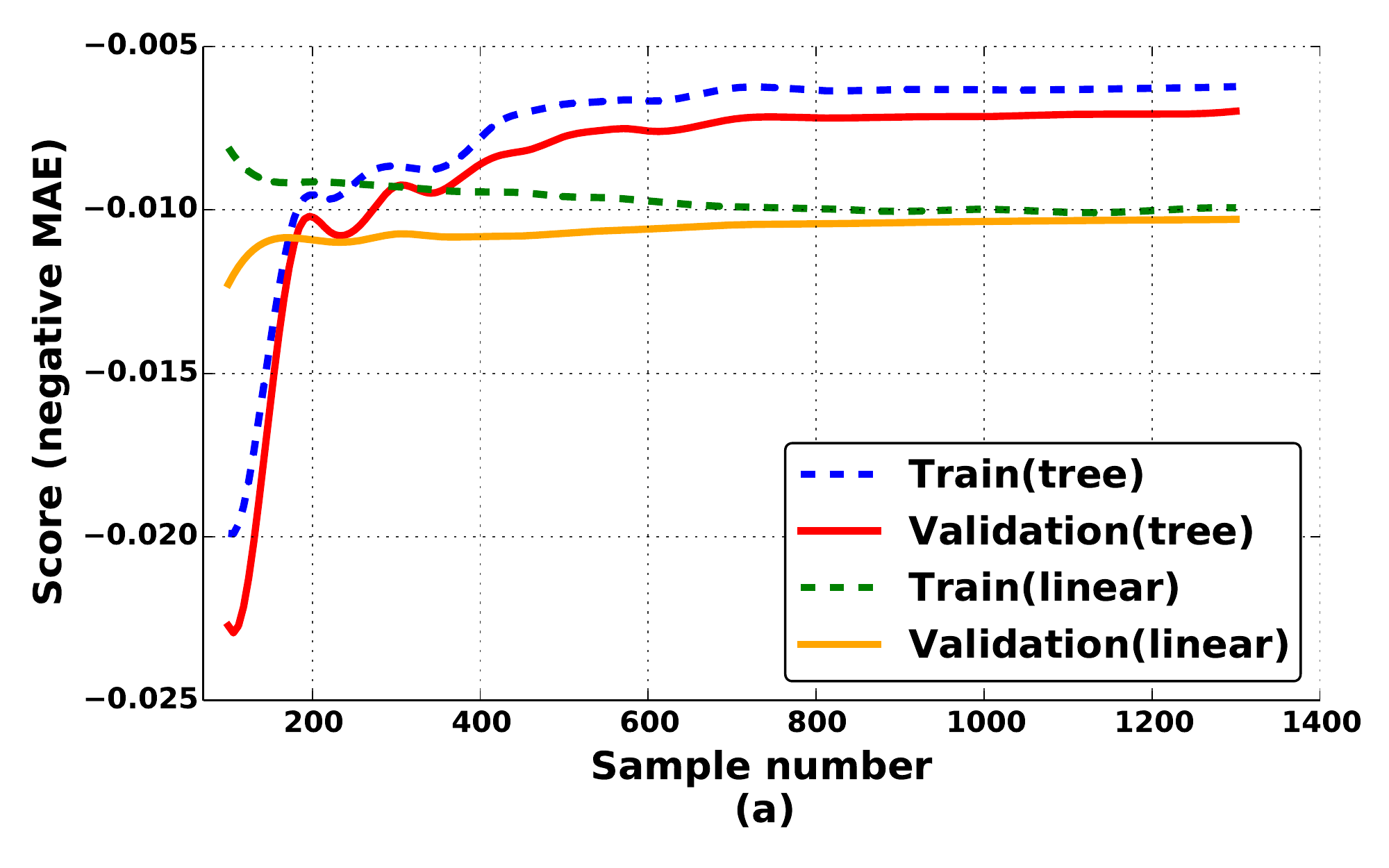}
\includegraphics[width=5.9cm]{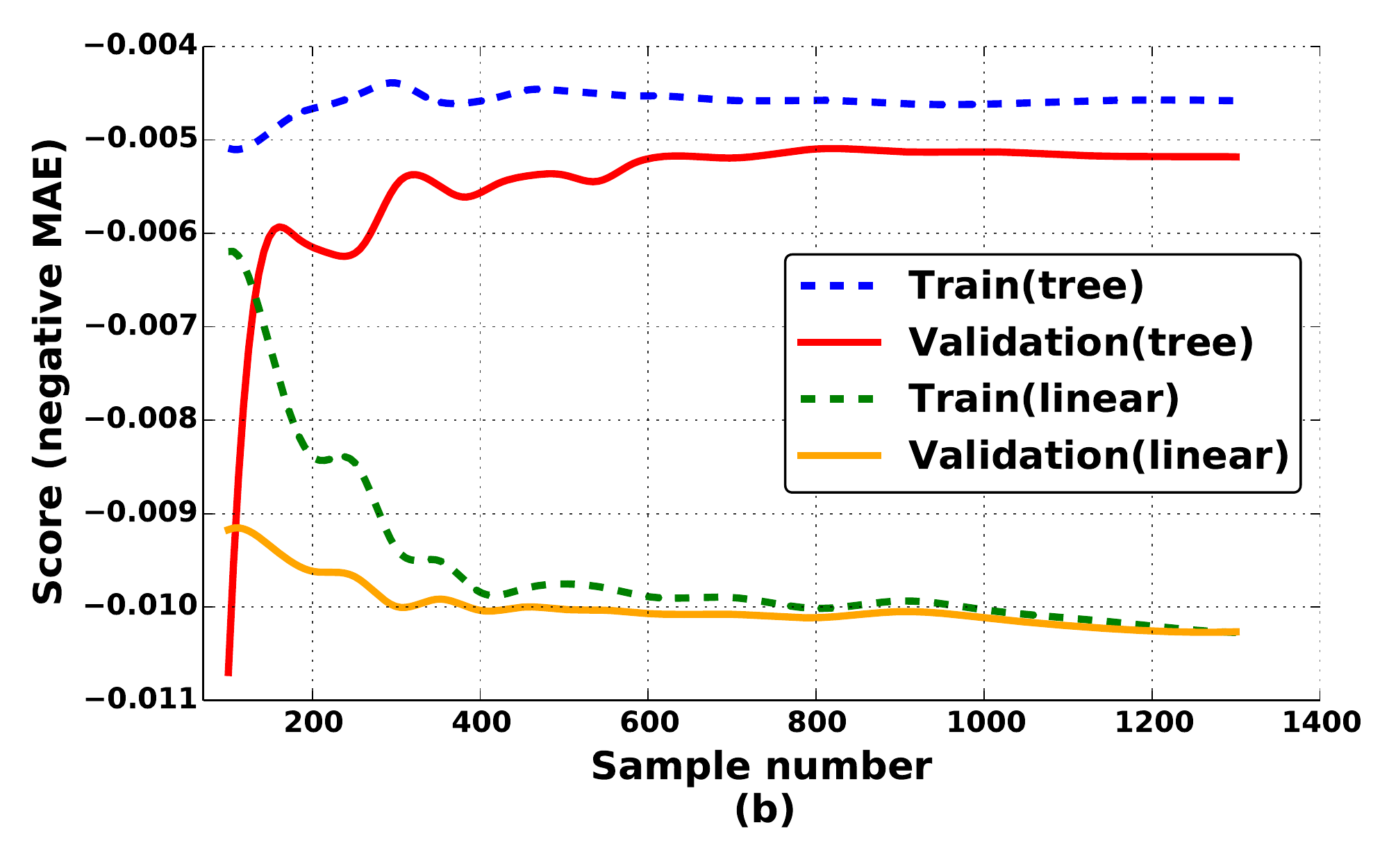}
\includegraphics[width=5.9cm]{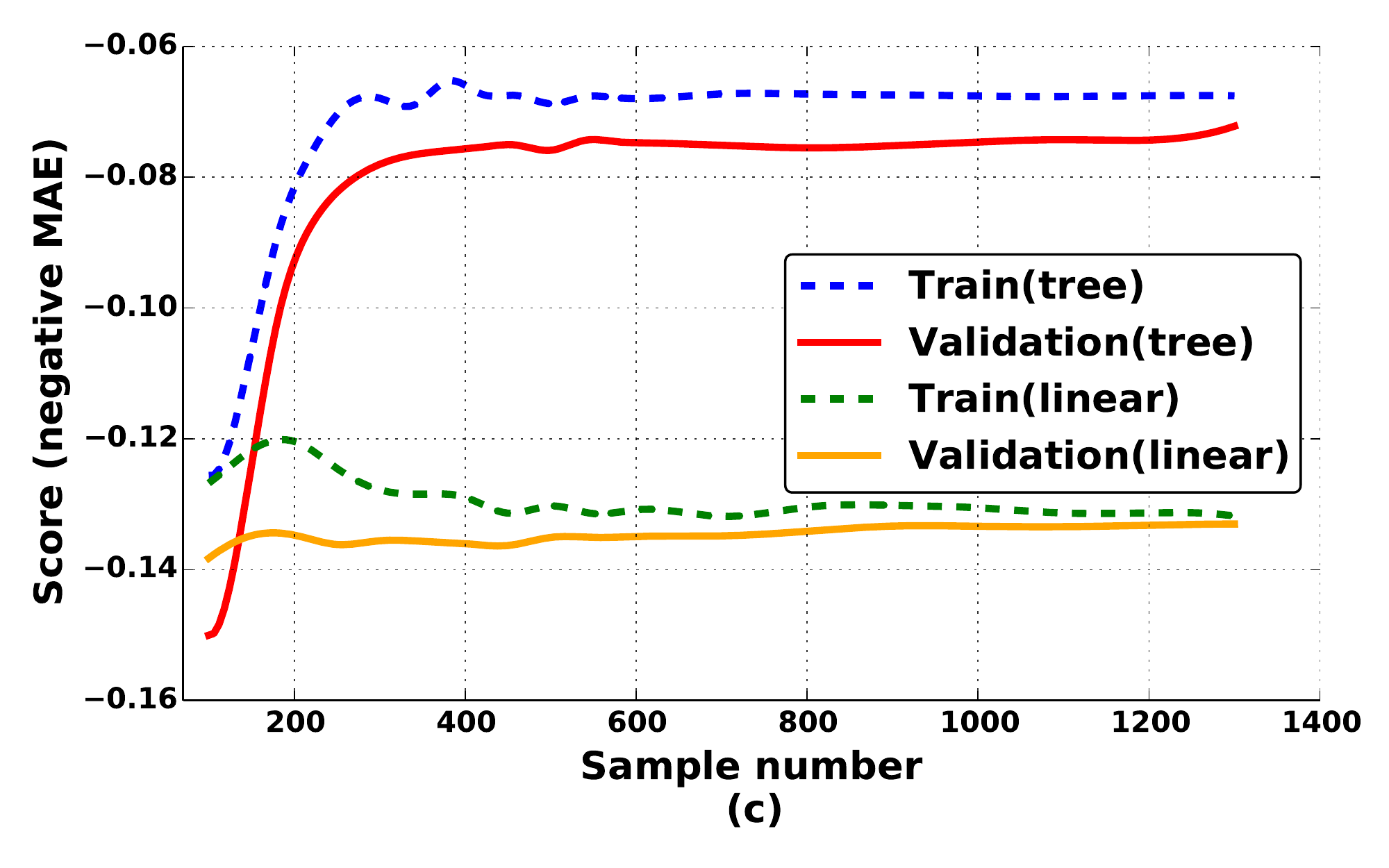}\\
\includegraphics[width=5.9cm]{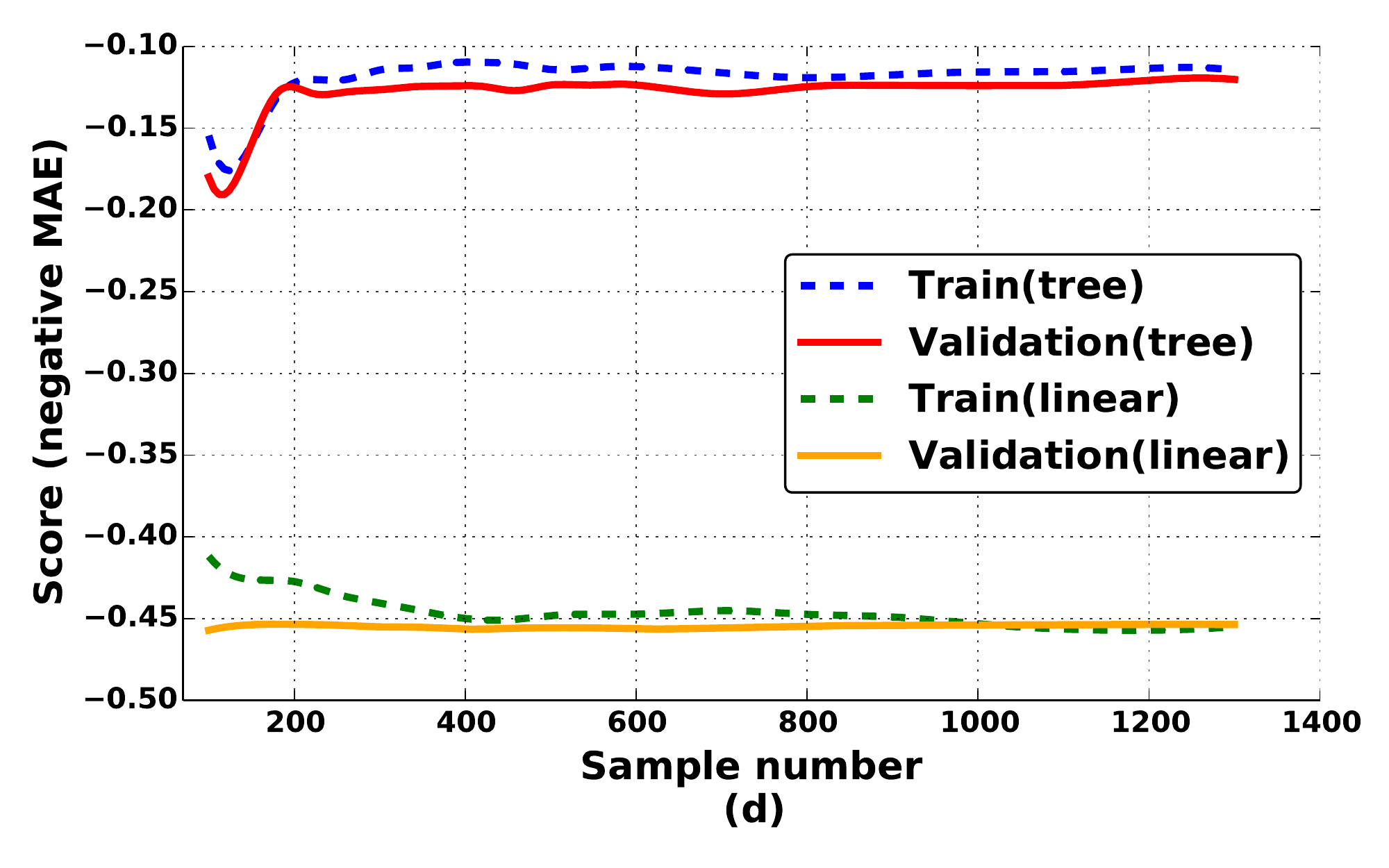}
\includegraphics[width=5.9cm]{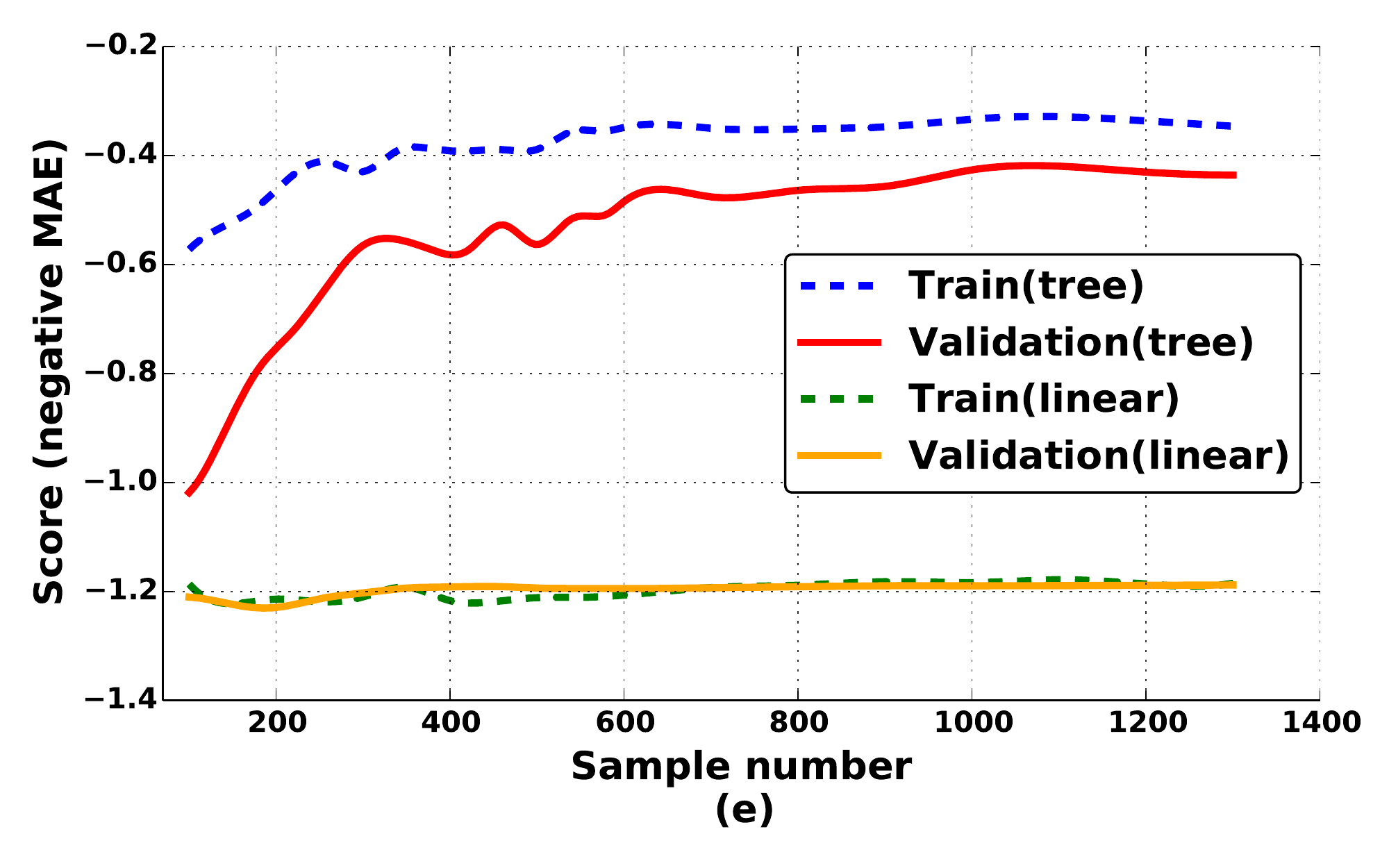}
\includegraphics[width=5.9cm]{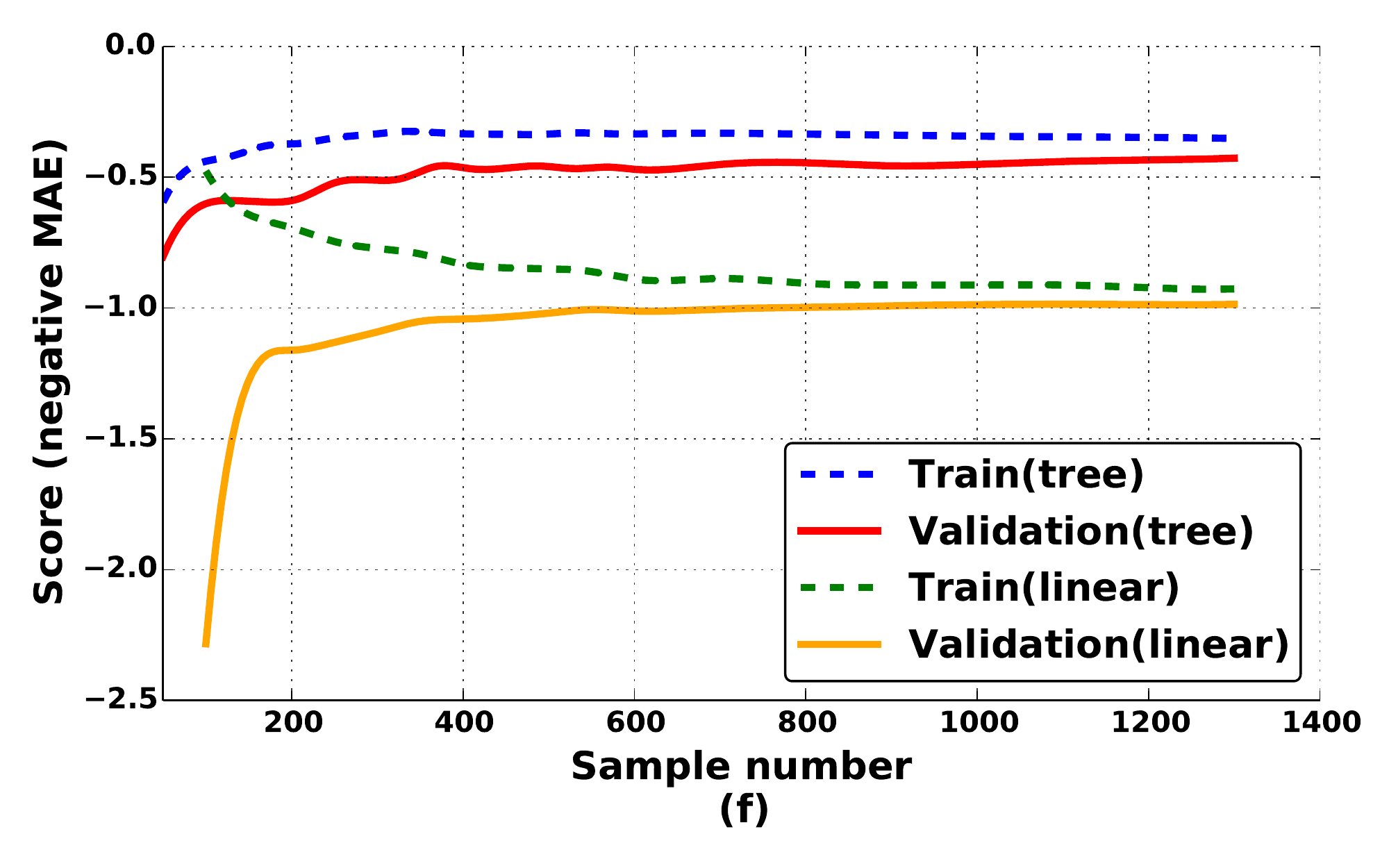}
\caption{Learning curve: (a) Atax; (b) Bicg; (c) GemmNcubed; (d) Matrixmult; (e) Hybrid\_1; (f) Hybrid\_2.}
\label{fig:learning}
\end{center}
\vspace{-4mm}
\end{figure*}
We systematically evaluate the model accuracy using different sets of benchmarks which can be categorized by the utilized resources. We define the following three types of benchmarks: LUT-based, DSP-based and hybrid. The LUT-based benchmarks (i.e., Atax and Bicg) mainly use LUT resources, whereas the DSP-based benchmarks (i.e., GemmNcubed and Matrixmult) utilize DSPs as the major resource type. The hybrid benchmarks are a combination of benchmarks showing a large proportion of resource utilization of both LUTs and DSPs. The LUT-based and DSP-based benchmarks are tested under the clock period of 10 ns and the hybrid benchmarks are evaluated with the clock period of 15 ns. The estimation period is 3$\mu$s. We collect 2000 samples for each benchmark by randomly invoking the applications with random input vectors. We select 20\% of the samples as the test set, whereas the other 80\% will be used to train the models as well as perform cross validation. We compare our decision-tree-based models with linear regression models employed by prior research work~\cite{laksh,najem,kapow}. In the feature selection step for linear models, we use the feature weights as the feature importance described in Section~\ref{subsec:fs} for a fair comparison.

The resource utilization and the mean absolute error (MAE) in percentage for dynamic power consumption is shown in Table~\ref{table:model}. The average MAE percentage is 4.36\% for our proposed decision tree model whereas the linear regression indicates an average error of 17.31\%. From Table~\ref{table:model}, we can see that the advantages of decision tree over linear model is larger for DSP-based and hybrid designs, because the LUTs are intrinsically better to be fitted in the linear regression model while the DSPs are inclined to have non-linear power patterns, as reported in~\cite{lee15,bogl} that the complex arithmetic units generally exhibit non-linear power behaviors.

The learning curves for training and cross validation further reveal the difference of the decision tree models and linear regression models regarding the capability to learn from samples, as shown in Fig.~\ref{fig:learning}. Note that the power consumption differs a lot for different benchmarks and thus the MAE varies largely for different benchmarks even with similar accuracy. Regarding the linear regression, the learning curves imply a high-bias scenario: the error is high and the model creases to improve accuracy given more training samples. In general, the high-bias situation means the underfitting problem of the training data has occurred. This also accounts for the deterioration in training accuracy as non-linear patterns increase with more samples. Comparatively speaking, the decision tree model exhibits more superior ability to learn from the training samples as the sample size gets larger. As a result, the decision trees exhibit notably lower errors compared with linear regression models.
\begin{table}
\begin{center}
\caption{Benchmark and Model assessment.}
\vspace{-2mm}
\label{table:model}
  \begin{tabular}[width=\linewidth]{c|c|c|c|c|c|c}
    \toprule
    \multirow{2}{*}{\textbf{Benchmark}} &
      \multicolumn{4}{c|}{\textbf{Resource (\%)}} &
      \multicolumn{2}{c}{\textbf{MAE (\%)}} \\
      & {LUT} & \multicolumn{1}{c|}{DSP} & \multicolumn{1}{c|}{FF} & {BRAM} & {Dtree} & {Linear} \\
      \midrule
    \multicolumn{1}{l|}{Atax} & 4.82 & 0 & 0.21 & \multicolumn{1}{c|}{0} & 4.14 & 12.46\\
    \multicolumn{1}{l|}{Bicg} & 2.72 & 0 & 0.14 & \multicolumn{1}{c|}{0} & 2.58 & 15.67\\
    \multicolumn{1}{l|}{GemmNcubed} & 1.16 & 53.33 & 1.46 & \multicolumn{1}{c|}{0} & 4.51 & 17.80\\
    \multicolumn{1}{l|}{Matrximult} & 6.24 & 100 & 3.65 & \multicolumn{1}{c|}{0} & 3.54 & 18.81\\
    \multicolumn{1}{l|}{Hybrid\_1}  & 53.82 & 97.78 & 20.04 & \multicolumn{1}{c|}{3.99} & 5.78 & 20.78\\
    \multicolumn{1}{l|}{Hybrid\_2} & 56.91 & 100 & 8.21 & \multicolumn{1}{c|}{5.69} & 5.61 & 18.34\\
    \bottomrule
  \end{tabular}
\vspace{-5mm}
\end{center}
\end{table}

\subsection{Frequency variation}
\begin{figure}[ht]
\begin{center}
\vspace{-3mm}
\includegraphics[width=.49\linewidth]{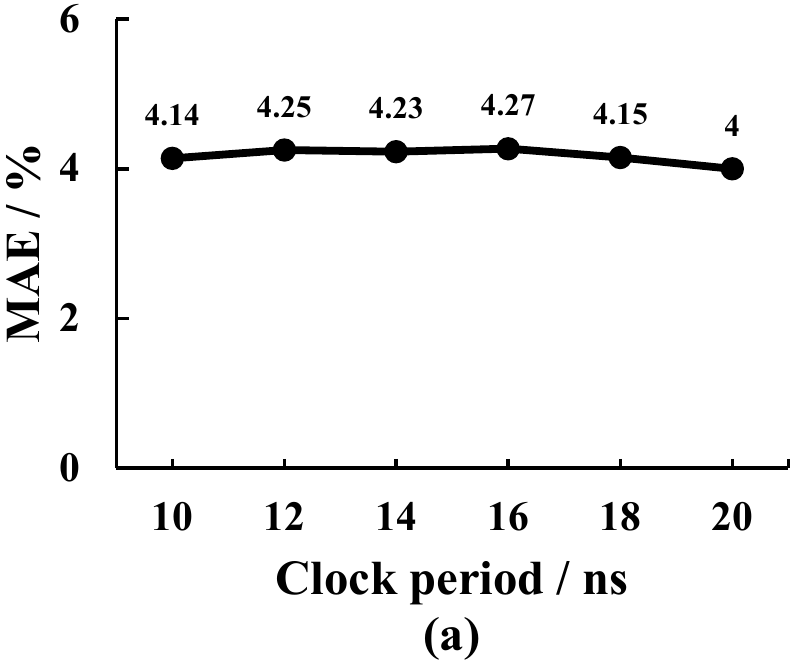}
\includegraphics[width=.49\linewidth]{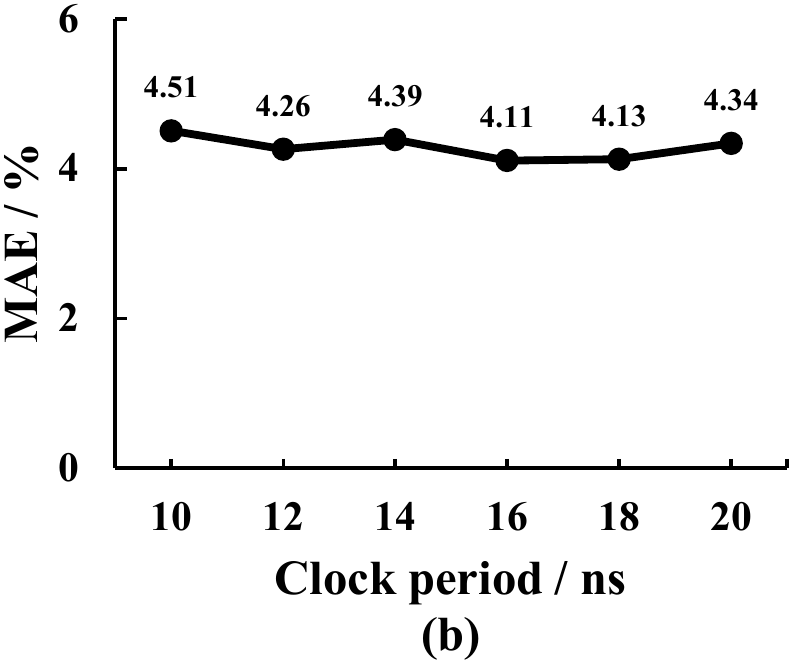}
\caption{Accuracy for frequency variation: (a) Atax (b) GemmNcubed.}
\label{fig:freq}
\vspace{-5mm}
\end{center}
\end{figure}

In this experiment, we evaluate our proposed model regarding the effect on estimation accuracy when the operating frequency varies. We use the pre-trained power models for Atax and GemmNcubed from Section~\ref{subsec:assess} to predict power consumption and verify the models' scalability when frequency changes at runtime. These models are trained under a clock period of 10 ns with the accuracy shown in Table~\ref{table:model}. We employ the same estimation cycles (i.e., 300 cycles) and run our design flow to collect new activity and power traces for testing purpose under different frequencies. Note that the power estimation works for the original period of 10 ns and we compute the predicted power by scaling the power estimation according to the ratio of frequencies (i.e., $\frac{f_{current}}{f_{model}}$) because the power consumption is proportional to the frequency as shown in Equation~(\ref{eq:dyn}). The results are shown in Fig.~\ref{fig:freq}. Compared with the baseline MAE for the clock period of 10 ns, the degradation of error is within 0.2\% under different frequencies. Thereby, our proposed model is applicable to be used under different operating frequencies with low generalization errors.

\subsection{Model overhead}
The decision tree hyper-parameter settings for the tested benchmarks are presented in Table~\ref{table:partun}. We analyze the extra overhead of integrating our proposed power models into the benchmarks from the following three aspects: resource utilization, operating frequency and power dissipation. The first three benchmarks utilize the DSP-based activity counters while the others use the LUT-based counterparts. The activity counter width can be tuned according to the estimation period as stated in Section~\ref{subsec:dtreereg}. Therein, we uniformly set the width as 20 bits which is sufficient to cover the highest activity for a wide range of estimation periods. For each single application, ten to twenty selected signals are monitored, each of which is equipped with an activity counter. As shown in Table~\ref{table:overhead}, the monitoring circuits consume less than 0.01\% of LUTs and FFs, 0.2\% of BRAMs and 0.4\% of DSPs. Note that by using LUT-based activity counters instead of DSP-based ones, we can further dispense with DSP resources. In contrast, the prior work using linear models~\cite{najem,kapow} demonstrate the LUT overheads of 7\% and 9\% respectively and the software overhead occupying 5\% CPU time. The decision tree exhibits much higher area-efficiency because it mainly leverages integer comparisons whereas the linear model occupies a large amount of floating-point additions and multiplications, which accounts for its high area overhead. The power dissipation of our decision tree model is extremely low. Besides this, the operating frequency fluctuates slightly, showing a maximum degradation of 0.70 MHz (1\%). In conclusion, our proposed decision-tree-based monitoring hardware demonstrates low overheads in resource utilization, operating frequency and power dissipation. It can be efficiently integrated in the RTL designs for on-board power monitoring.

\begin{table}[ht]
\vspace{-3mm}
\caption{Decision tree hyper-parameter settings.}
\begin{center}
\begin{tabular}[width=\linewidth]{ccccc}
    \toprule
    \multicolumn{1}{c|}{\multirow{2}{*}{\textbf{Benchmark}}} & \multicolumn{1}{c|}{\multirow{1}{*}{\textbf{Max}}} & \multicolumn{1}{c|}{\multirow{1}{*}{\textbf{Min split}}} & \multicolumn{1}{c|}{\multirow{1}{*}{\textbf{Min leaf}}} & \multicolumn{1}{c}{\multirow{1}{*}{\textbf{Min leaf}}}\\
     \multicolumn{1}{c|}{} & \multicolumn{1}{c|}{\textbf{depth}}& \multicolumn{1}{c|}{\textbf{sample}}& \multicolumn{1}{c|}{\textbf{sample}}& \multicolumn{1}{c}{\textbf{impurity}} \\ \midrule

   \multicolumn{1}{l|}{Atax}&
   \multicolumn{1}{c|}{5}&\multicolumn{1}{c|}{20}&\multicolumn{1}{c|}{20}&\multicolumn{1}{c}{0.01}\\

   \multicolumn{1}{l|}{Bicg}&
   \multicolumn{1}{c|}{5}&\multicolumn{1}{c|}{5}&\multicolumn{1}{c|}{5}&\multicolumn{1}{c}{0.001}\\

   \multicolumn{1}{l|}{GemmNcubed}&
   \multicolumn{1}{c|}{5}&\multicolumn{1}{c|}{20}&\multicolumn{1}{c|}{20}&\multicolumn{1}{c}{0.03}\\

   \multicolumn{1}{l|}{Matrixmult}&
   \multicolumn{1}{c|}{4}&\multicolumn{1}{c|}{5}&\multicolumn{1}{c|}{5}&\multicolumn{1}{c}{0.03}\\

   \multicolumn{1}{l|}{Hybrid\_1}&
   \multicolumn{1}{c|}{6}&\multicolumn{1}{c|}{5}&\multicolumn{1}{c|}{5}&\multicolumn{1}{c}{0.02}\\

   \multicolumn{1}{l|}{Hybrid\_2}&
   \multicolumn{1}{c|}{6}&\multicolumn{1}{c|}{20}&\multicolumn{1}{c|}{10}&\multicolumn{1}{c}{0.05}\\
     \bottomrule
\end{tabular}
\vspace{-5mm}
\end{center}
\label{table:partun}
\end{table}

\begin{table}[ht]
\caption{Model overhead analysis.}
\begin{center}
\begin{tabular}[width=\linewidth]{cccccccc}
    \toprule
    \multicolumn{1}{c|}{\multirow{2}{*}{\textbf{Benchmark}}} & \multicolumn{4}{c|}{\multirow{1}{*}{\textbf{Resource} (in number)}} & \multicolumn{1}{c|}{\multirow{1}{*}{\textbf{Freq}}} & \multicolumn{1}{c}{\multirow{1}{*}{\textbf{Power}}}\\
     \multicolumn{1}{c|}{} & \multicolumn{1}{c|}{LUT}& \multicolumn{1}{c|}{DSP}& \multicolumn{1}{c|}{FF}& \multicolumn{1}{c|}{BRAM} & \multicolumn{1}{c|}{(MHz)} & \multicolumn{1}{c}{(mW)} \\ \midrule

   \multicolumn{1}{l|}{Atax}&
   \multicolumn{1}{c|}{127}&\multicolumn{1}{c|}{7}&\multicolumn{1}{c|}{198}&\multicolumn{1}{c|}{2.5}&
   \multicolumn{1}{c|}{+1.14}&
   \multicolumn{1}{c}{3}\\

   \multicolumn{1}{l|}{Bicg}&
   \multicolumn{1}{c|}{125}&\multicolumn{1}{c|}{6}&\multicolumn{1}{c|}{176}&\multicolumn{1}{c|}{0.5}&
   \multicolumn{1}{c|}{+2.32}&
   \multicolumn{1}{c}{2}\\

   \multicolumn{1}{l|}{GemmNcubed}&
   \multicolumn{1}{c|}{149}&\multicolumn{1}{c|}{9}&\multicolumn{1}{c|}{242}&\multicolumn{1}{c|}{2.5}&
   \multicolumn{1}{c|}{+1.35}&
   \multicolumn{1}{c}{4}\\

   \multicolumn{1}{l|}{Matrixmult}&
   \multicolumn{1}{c|}{108}&\multicolumn{1}{c|}{0}&\multicolumn{1}{c|}{325}&\multicolumn{1}{c|}{0.5}&
   \multicolumn{1}{c|}{0}&
   \multicolumn{1}{c}{4}\\

   \multicolumn{1}{l|}{Hybrid\_1}&
   \multicolumn{1}{c|}{156}&\multicolumn{1}{c|}{0}&\multicolumn{1}{c|}{415}&\multicolumn{1}{c|}{2.5}&
   \multicolumn{1}{c|}{-0.70}&
   \multicolumn{1}{c}{5}\\

   \multicolumn{1}{l|}{Hybrid\_2}&
   \multicolumn{1}{c|}{162}&\multicolumn{1}{c|}{0}&\multicolumn{1}{c|}{508}&\multicolumn{1}{c|}{2.5}&
   \multicolumn{1}{c|}{+1.62}&
   \multicolumn{1}{c}{6}\\
     \bottomrule
\end{tabular}
\vspace{-6mm}
\end{center}
\label{table:overhead}
\end{table}

\subsection{Model ensemble}
In order to verify the accuracy of the ensemble strategy, we combine Atax, Bicg and GemmNcubed with the pre-trained power models and run the design flow to generate testing samples. We then quantify the accuracy of aggregating separately pre-trained models as an ensemble model by adding up the predicted power for different components. As a comparison, we re-train a monolithic power model using new samples. The MAE percentage of the ensemble model and the re-trained model is 5.52\% and 4.32\%, respectively. The accuracy deterioration of the ensemble model is 1.2\%. This is mainly owing to the changes in placement and routing. In all, the model ensemble strategy dispenses with the need for re-training new models with moderate accuracy degradation.

\subsection{Fine-grained phase shedding for on-chip multi-phase voltage regulator}
Noticing that the phase shedding for on-chip regulator shows prominent efficiency improvement for the processor in~\cite{vrscale} by reducing the number of phases in light load situations, we similarly investigate the viability for phase shedding in FPGA for internal logic supplied by $V_{ccint}$. However, different from processors which internally incorporate multiple power states, applications running in FPGA are fully customized by designers and intrinsically there is a lack of indicative states about runtime power to guide the phase shedding decision. Following this observation, we leverage our proposed power estimator to supervise a fine-grained phase shedding strategy of an on-chip regulator.

We synthesize a two-stage power delivery system consisting of both on-chip and off-chip multi-phase voltage regulators for $V_{ccint}$ using the tool PowerSoC~\cite{wang15}. Both the on-chip and off-chip regulators are buck converters with the parameters specified in~\cite{wang15}. To approximate the power loss induced by off-chip to on-chip and on-chip to die parasitic resistances, we estimate the fabrication details including the resistance of package vias and wires, interposer vias and wires, micro-bumps and TSVs from work~\cite{xu12,wrona,chaware}. The number of the above network components for the targeted device are estimated from~\cite{chaware,sp7}. Finally the power delivery system for internal logic is constructed under the nominal power of 20 W. The on-chip regulator has five phases with the voltage scaling time between $V_{ccint}$ and ground within 20ns, which is safe for fine-grained controlling with our 3$\mu$s power estimation period.

\begin{figure}[ht]
\begin{center}
\vspace{-3mm}
\includegraphics[width=\linewidth]{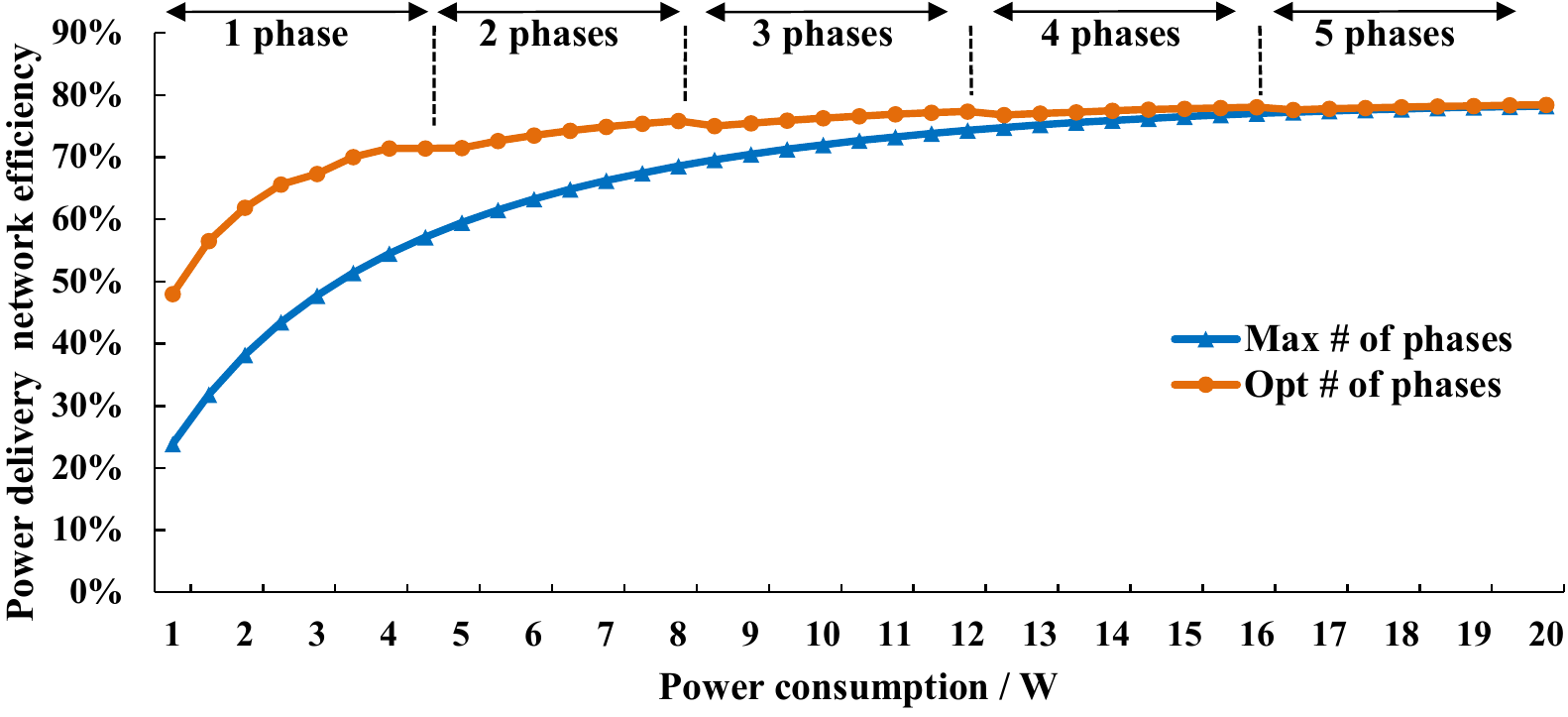}
\vspace{-7mm}
\caption{Efficiency of power delivery network for FPGA internal logic.}
\label{fig:vr}
\vspace{-3mm}
\end{center}
\end{figure}

We first experimentally determine the optimal number of phases that can maximize the power efficiency of the power delivery network under different power values as shown in Fig.~\ref{fig:vr}. Then, we deploy a look up table based phase shedding approach~\cite{vrscale} and determine the optimal number of phases to use at runtime based on the power of internal logic including static power and monitored dynamic power. The power efficiency improvement is derived according to Equation~(\ref{eq:vreff}) where $i$ denotes the index of  a specific estimation period, $P_{nopt}$ and $P_{nmax}$ are the power under optimal number of phases and maximum number of phases, respectively. The phase shedding induced power overhead $P_{loss}$ is calculated according to~\cite{wang15}.

Experiments show that the efficiency improvement for Hybrid\_1 and Hybrid\_2 using the last 400 sample sequences from Section~\ref{subsec:assess} are 13.6\% and 14.4\% respectively, with the internal logic power ranging from 1 W to 11 W. As a result, the proposed runtime power monitoring hardware provides fine-grained power information for runtime phase shedding of an on-chip regulator to boost 14\% average efficiency of the power delivery network. It is promising to be used in future FPGAs with integrated voltage regulators and CPU-FPGA system-on-chips as well for further power saving.

\begin{equation}
\label{eq:vreff}
Eff_{impv}=1-\dfrac{\sum_i^{N} (P_{nopt}(i)+P_{loss}(i))}{\sum_i^NP_{nmax}(i)}
\end{equation}

\section{conclusion}
\label{sec:con}
In this work, we leverage the state-of-the-art machine learning techniques to establish a novel decision-tree-based dynamic power monitoring approach for applications running on FPGA. The proposed design flow and the decision tree regression model can be generally applied for fine-grained power prediction in FPGA. We also develop a light-weight and in-situ monitoring hardware for the developed power model which can be efficiently integrated into RTL designs with extremely low overheads of area, power and performance. We investigate our proposed methodology on three different types of benchmarks: LUT-based, DSP-based and hybrid benchmarks. Experimental results reveal that the decision tree models outperforms the traditional linear models with more than 10\% reduction in mean absolute error. Besides this, the decision-tree-based power monitoring exhibits a high capability to learn from samples as the training sample size increases whereas the linear regression is prone to the underfitting problem. Moreover, we exploit a model ensemble method used for library-based and IP-based designs, with the results exhibiting an additional 1.2\% error compared with a completely re-trained model. Furthermore, we utilize our proposed power monitoring scheme to guide the phase shedding of an on-chip multi-phase regulator as a case study and the results demonstrate 14\% average improvement in the efficiency of the power supply for the FPGA internal logic. In future work, we plan to enhance the power monitoring scheme with static power compensation and leverage fine-grained runtime power monitoring for advanced power management schemes.

\balance
\bibliographystyle{IEEEtran}
\bibliography{Ref}

\end{document}